\documentstyle[sprocl]{article}

\def\be{\begin{equation}}
\def\ee{\end{equation}}
\def\bea{\begin{eqnarray}}
\def\eea{\end{eqnarray}}

\def\Tr{\rm Tr~}

\begin{document}
\title{Mass Hierarchies from Anomalies: \\a Peek\\ Behind the 
Planck Curtain}
\author{P. Ramond}
\address{Institute for Fundamental Theory, Physics Department.
\\ University of Florida, Gainesville, FL 32611}

\maketitle\abstracts{The masses of quarks and leptons suggest a 
strong hierarchical structure. We argue that their patterns  can be 
reproduced through the introduction of a new  Abelian symmetry. 
The data suggest that this symmetry is anomalous. We suggest 
that the cancellation of its anomalies occur through the 
Green-Schwarz mechanism. An important check of this idea is 
that it links the Weinberg angle to a mass ratio of the 
elementary fermions. The Green-Schwarz mechanism 
occurs naturally in many superstring compactifications, and 
produces a small parameter, which we use to determine the quark mass 
hierarchy.  We show that hierarchy and mixings among the 
chiral fermions is a consequence of the Green-Schwarz 
mechanism. We  present several models where this idea is 
realized}  

\section{Introduction} Superstring and related theories of extended 
objects, to which Keiji Kikkawa has contributed so much, offer 
the best hope for resolving the difference between General 
Relativity and Quantum Mechanics. Remarkably they  contain the gauge 
structures found in the description of low energy phenomena.  
Yet, in spite of this conceptual matching between the 
$``$observed" and the $``$fundamental", the lack of detailed 
predictions, such as relations among the parameters and structures 
of the standard model, has undermined the credibility of these 
theories as candidate Theories of Nature. The disparity of 
scales has made it difficult to equate the two: the standard 
model is an effective theory with a cut-off somewhere between 1 
TeV and the Planck mass;  superstring theories apply to physics 
above the Planck mass. However, as shown by 't Hooft, anomalies are 
impervious to scales. For example in QCD, the global chiral anomaly 
responsible for the decay of the $\pi^0$ $``$measures" the 
number of colors, at a scale well below that at which quarks 
become apparent. 

As a chiral theory, the standard model could have many 
anomalies, but as is well known the hypercharge anomalies
cancel between quarks and leptons. Curiously the 
mixed anomaly between the hypercharge current and the 
energy-momentum tensors  cancels as well. Hence to use  
 anomalies as probes of fundamental physics,  we must 
extend the standard model.

Below we argue that the puzzling pattern of quark and lepton 
masses can be generated by adding an Abelian symmetry to the 
standard model, following the idea of Froggatt and 
Nielsen~\cite{FN}. 
Remarkably, their  simple picture can reproduce the features of the CKM matrix, 
and  the mass hierarchies, if the Abelian symmetry is anomalous~\cite{BR,BLR}. 

Many superstring compactifications contain in their low energy manifestation,  
a chiral  Abelian symmetry, with anomalies $``$cancelled" by the Green-Schwarz 
mechanism~\cite{GS}. This symmetry is broken spontaneously {\em below} the string cut-off, 
automatically producing a small parameter that can be used to generate the mass 
hierarchies. The Weinberg angle at the string cut-off is then 
determined~\cite{Ib} in terms 
of the anomaly coefficients, which depend on the charges of the massless particles.

To match the two theories through this $U(1)$~\cite{IR,BR},  their 
cut-offs must be at the same scale. For the standard model, this requires its extension by low 
energy supersymmetry. Thus all our remarks below will be made in that context.

\section{ The Yukawa Sector}

The  regularities found in the gauge sector of the 
standard model, do not seem to have any counterparts in the 
Yukawa sector, the most studied, least understood, and 
therefore the most interesting sector of the standard model. 

The high value of the mass of the top quark~\cite{top} enables a very 
attractive mechanism for electroweak breaking, triggered through the 
renormalization group by   soft 
supersymmetry breaking terms. This picture predicts the existence of 
the superpartners of the elementary  particles with masses of the order 
of hundred(s) of GeVs. The same theory also allows us to extrapolate the 
physical parameters of the standard model to its supersymmetric cut-off, 
as the theory is perturbative throughout that range. 

The Yukawa sector contains many small parameters, such as the ratio of the 
bottom to top quark masses, the ratio of electron to muon masses, 
etc..., as well as small mixing angles. A convenient parametrization~\cite{RRR} 
of these mass parameters, introduced for the CKM matrix by 
Wolfenstein~\cite{wolf}, 
uses the Cabibbo angle $\lambda=V^{}_{us}$ as the small parameter. 
If we assume no 
significant matter between the TeV scale and the cut-off, as implied 
by the convergence of the gauge couplings, we find the following 
ratios, valid around $10^{16}$ GeV, slightly below the string 
cut-off:
\begin{equation}
{m_u\over m_t}={\cal O}(\lambda^8)\ ;\qquad {m_c\over m_t}={\cal
O}(\lambda^4)\ ; \label{eq:lag} \end{equation}
\begin{equation}{m_d\over m_b}={\cal
O}(\lambda^4)\ ; \qquad {m_s\over m_b}={\cal
O}(\lambda^2)\ ,\label{eq:l} 
\end{equation}
The charged lepton masses satisfy similar relations
\begin{equation}
{m_e\over m_\tau}={\cal
O}(\lambda^4)\ ; \qquad {m_\mu\over m_\tau}={\cal
O}(\lambda^2)\ .\label{eq:la} 
\end{equation}
Finally, we have the interfamily hierarchy
\begin{equation}
{m_b\over m_t}={\cal O}(\lambda^3)\ .\label{eq:lb} 
\end{equation}
The Froggatt and Nielsen~\cite{FN} idea relates the  exponents that appear in 
the mass ratios are related to  the dimension of the 
operators that generated them. 
In the following we present a general framework to implement this idea, 
and then present  several models, using invariance under the anomalous 
$U(1)_X$ and under other possible Abelian symmetries present in 
superstring-generated theories.

\section{Profile of Superstring-Generated Effective Theories}

At a time when dualities suggest that  all  superstring theories are 
manisfestations of one  unknown theory,  we still do not know the reason nor the 
dynamical mechanism by which they compactify to  four space-time dimensions. Such 
ignorance forces us to consider   compactifications on all types of manifolds, 
generating a large number of  candidate four dimensional theories. 

While we cannot put much credence on the detailed predictions of any one of these 
compactifications, many contain similar characteristic  features.  In that spirit, 
we study a class of compactifications~\cite{faraggi} which contains  a gauged phase symmetry, 
which has anomalies cancelled in a way specific to superstring theories, first 
discovered by Green and Schwarz~\cite{GS}.

Compactified superstring theories engender at and below a 
string cut-off, $M_{string}$ an effective $``$low energy" 
theory. This theory  is a gauge theory, generically invariant 
under $N=1$ supersymmetry, with  one universal gauge coupling 
$\alpha_{string}$. These two parameters are related by the 
Planck mass.  The gauge structure of these theories is 
separated in the $``$visible" and $``$hidden" sectors, 
connected to one another by an $``$anomalous" phase symmetry, 
$U(1)_X$, and possible other straddling but non-anomalous phase symmetries: 
\begin{equation}
{\cal G}^{visible}\times U(1)_X\times 
U(1)\times\cdots\times U(1)\times {\cal G}^{hidden}\ .
\label{eq:dsw} 
\end{equation} 
The first gauge group describes the $``$visible sector", with gauge 
group which may be a GUT group, but must at least contain the 
standard model structure 
\begin{equation}
{\cal 
G}^{visible}\subset SU(3)^C\times SU(2)^W\times 
U(1)^Y\times\cdots \ .
\label{eq:gvis} 
\end{equation}
 In superstring compactifications, the 
dots stand for Abelian symmetries. The  gauge group of the 
hidden sector is ${\cal G}^{hidden}$. Its precise form  is 
model dependent. The matter chiral superfields which transform 
under ${\cal G}^{hidden}$ are singlets under ${\cal 
G}^{visible}$, and vice-versa. The two sectors 
interact through the universal gravitational interactions, and  
a set of Abelian forces. The interactions of the hidden sectors 
are assumed to be strong, generating perhaps dynamical breaking 
of supersymmetry by means of gaugino condensation. 

When viewed below the cut-off of an effective field theory, the straddling 
symmetry, $U(1)_X$, is anomalous: the massless fermions  produce anomalous 
divergences of its current, to be compensated by  cut-off dependent terms. For 
superstrings, this term is an axion-like coupling term of dimension five. This 
mechanism is the four-dimensional analogue of the Green-Schwarz anomaly 
cancellation of the mother superstring theory in ten dimensions. The theory may 
contain other straddling $U(1)$s, they  are not anomalous, although model dependent.

In addition, the anomalous $U(1)_X$ is spontaneously broken by a stringy mechanism 
a few orders of magnitude {\it below} the string cut-off. This mechanism, first 
noticed by Dine, Seiberg, and Witten~\cite{DSW}, generates a one-loop finite 
contribution to its D-term

\begin{equation}
D^{}_X=D_X^{tree}+{g^3_{string}\over 
192\pi^{2}}~M^2_{string}C^{}_{\rm grav}\ ,\label{eq:dsw2} 
\end{equation}
where $D_{X}^{tree}$ is the tree level D term, and $C^{}_{\rm grav}$ is the sum 
over helicities of the $X$-charges of all the fermions, the mixed gravitational 
anomaly of the $U(1)_X$ charge
\be 
C^{}_{\rm grav}=\Tr(XTT)\ ,\label{eq:gs0}
\ee
where $T$ stands for the energy momentum tensor. As  the breaking 
occurs below the cut-off, the effective low energy theory is invariant 
under $U(1)_X$, imposing constraints on the form of the low energy 
theory, and generating a small parameter, the 
ratio of the breaking scale to the string cut-off. Below, 
we argue that this small parameter is at the origin of the fermion 
mass hierarchies. 

Let us specialize these equations to theories that contain the 
standard model. We take the non-Abelian visible gauge group to 
be ${\cal 
G}^{visible}_{}=SU(3)^C\times SU(2)^W\times U(1)_1\times U(1)_2\times
\cdots\times U(1)_K$.
The hypercharge $Y$ is a linear combination of these Abelian 
charges,   generated by currents $Y^{}_k$,  $k=1,2\dots, K$. 
 
We denote the mixed anomaly between the $X$ 
current and the non-Abelian gauge currents by $C^{}_{\rm color}$ 
and $C^{}_{\rm weak}$

\be  \Tr(XG^AG^B)=\delta^{AB}C^{}_{color}\ ;\qquad 
\Tr(XW^aW^b)=\delta^{ab}C^{}_{weak}\ ,\label{eq:gs1}\ee
where $G^A$ are the QCD currents, and $W^a$ the weak isospin 
currents. The Green-Schwarz mechanism requires  

\be \Tr (XY^{}_iY^{}_j)=\delta^{}_{ij}C^{}_i\ .\label{eq:gs2}\ee
All the other anomaly coefficients must vanish by themselves
\be \Tr(Y^{}_iY^{}_jY^{}_k)=\Tr(Y^{}_iXX)=0\ .\label{eq:gs3}\ee

The Green-Schwarz anomaly cancellation mechanism is understood in the 
language of an effective theory by writing its Lagrangian as

\begin{equation}{\cal L}={1\over g^2_{string}}\sum_{j} 
k^{}_{j}F^{[j]}_{\mu\nu}
F^{[j]}_{\mu\nu}+i{\eta\over M^{}_{string}}\sum_j k^{}_jF^{[j]}_{\mu\nu}
{\tilde F}^{[j]}_{\mu\nu}+{\cal L}^{}_{matter}+\cdots\ 
,\label{eq:lc} \end{equation}
where the $k^{}_j$ are the Kac-Moody levels, the index $j=$color, 
weak, $k$,...,hidden, denotes all 
possible gauge groups of the model, with field strengths 
$F^{[j]}_{\mu\nu}$, and $\eta$ is the axion, part of the dilaton 
supermultiplet, the transverse remnant of the Kalb-Ramond 
antisymmetric tensor field. 
The axion also couples to the gravitational field through a term 
symbolically denoted by $k^{}_g\eta R\tilde R$. 
The divergence of the $U(1)_X$ current is altered by the anomalies
\begin{equation}\partial^{}_\mu J^X_\mu\sim \sum_{j} C^{}_jF^{[j]}_{\mu\nu}
{\tilde F}^{[j]}_{\mu\nu}+C^{}_{\rm grav}R\tilde R\ ,\label{eq:div} 
\end{equation}
including for completeness the gravitational contribution to 
the anomaly. 

Under a gauge transformation of $U(1)_X$, the Lagrangian changes by an 
amount proportional to the divergence of this current, thereby 
creating an apparent lack of invariance. However, the axion field 
shifts at the same time, which generates a term of exactly the same 
form as that from the divergence of the current, provided that 
\begin{equation}
{C^{}_{\rm grav}\over 12}={C^{}_{\rm color}\over 
k^{}_{\rm color}}={C^{}_{\rm weak}\over k^{}_{\rm weak}}
=\cdots={C^{}_{i}\over k^{}_{i}}=\cdots\ne 0\ .\label{eq:gs} 
\end{equation}
The normalization between the mixed gravitational and mixed gauge 
anomalies is fixed in superstring theories. In any superstring derived
theory, these equations are always satisfied since the underlying
theory is anomaly-free. It is just that an apparent anomaly has 
been introduced by the cut-off procedure.

One can show~\cite{ir} that  the 
Green-Schwarz mechanism is unaffected by a canonical 
change of basis among the Abelian factors, $Y_i\rightarrow 
\tilde Y_i$ , that is
\be 
\Tr(X\tilde Y_i\tilde Y_j)=\delta_{ij}\tilde C_i\qquad 
{\tilde C_i\over \tilde k_i}={C_i\over k_i}\ .\label{eq:canon}
\ee
It follows that we can always choose the hypercharge as one of 
the $Y_i$'s, say $Y_1$, and drop the tildes.

The implications of these equalities are far reaching. As  
noted by Ib\'a\~nez~\cite{Ib},  they can be used to relate 
the Weinberg angle at the cut-off to 
the anomaly coefficients. Indeed, we have 
\begin{equation}
\tan^2\theta^{}_W={g_Y^2\over g_{\rm weak}^2}=
{k^{}_Y\over k^{}_{\rm weak}}={C^{}_Y\over C^{}_{\rm weak}}\ .
\label{eq:iba} \end{equation}
All the Kac-Moody levels are the same for the non-Abelian groups. 

The second consequence of these equations is that the mixed 
anomaly with the hidden sector gauge groups do not vanish, 
implying the existence of massless particles with hidden 
non-Abelian charges. If none are to survive as 
massless particles, they must all be confined, implying the 
existence of strong hidden forces, which might break 
supersymmetry dynamically. It also means 
that the mixed gravitational anomaly contains contributions 
from hidden sector particles. 

\section{ Anomaly Constraints on the Visible Sector}

Consider the minimal supersymmetric standard model with  three chiral 
families.  Out of its chiral supermultiplets, it is possible to 
form analytic combinations that are both electroweak and 
$R$-invariants:
 
\begin{equation}
\begin{array}{rcl}
{\cal O}^{[u]}_{ij}&=&{\bf Q}^{}_i{\bf\overline u}^{}_jH^{}_u\ ;
\qquad {\cal O}^{[d]}_{ij}={\bf Q}^{}_i{\bf\overline d}^{}_jH^{}_d
\ ;\\
{\cal O}^{[e]}_{ij}&=&L^{}_i{\overline e}^{}_jH^{}_d\ ;\qquad 
 {\cal O}^{[\mu]}_{}=H^{}_uH^{}_d\ , 
\\ & & \end{array}\label{eq:smo} \end{equation}
where $i,j$ are the family indices. 
With one right-handed neutrino for each chiral family, the list 
includes more terms
\begin{equation}
{\cal O}^{[\nu]}_{ij}=L^{}_i{\overline N}^{}_jH^{}_u\ ;\qquad
{\cal O}^{[0]}_{ij}={\overline N}^{}_i{\overline N}^{}_j\ 
.\label{eq:neut} 
\end{equation}

In the visible sector, there could be other chiral multiplets. So as not 
to contradict data, they should appear as vector-like pairs under the 
standard model gauge groups, although they could be chiral with respect 
to symmetries beyond. Their presence would not affect $C^{}_{\rm color}$, 
$C^{}_{\rm weak}$, nor $C^{}_Y$, but would affect $C^{}_k$.  

These combinations are the lowest dimension invariants. Without  any 
symmetry beyond the minimal supersymmetric standard model, all can appear 
in the superpotential, providing no explanation for any hierarchy among 
the Yukawa couplings.

As we have just discussed, low energy theories derived from superstrings 
can be expected to contain extra Abelian symmetries. The above  
invariants that do not respect these symmetries may still appear as 
higher dimension operators in the effective superpotential, dressed with 
new operators required by  the Abelian invariances, but suppressed by 
inverse powers of the cut-off. This will produce, upon the breaking of 
the phase symmetries, in a hierarchical structure of Yukawa couplings. 
Remarkably, it turns out that a hierarchy is generated if one of these 
extra charges is the Green-Schwarz symmetry. 

The charges of the chiral superfields, with family index 
$i=1,2,3$ are summarized in the following table, where we have 
separated the hypercharge from the other non-anomalous $U(1)$'s.
\vspace{0.4cm}
\begin{center}
\begin{tabular}{|c|c|c|c|c|c|c|c|c|c|}
\hline & & & & & & & & &  \\
$~{\rm Charge}~$&$~{\bf Q}_i~$&$\overline{\bf u}_i~$&
$~\overline{\bf d}_i~$&$~L_i~$&$~\overline e_i~$&$~\overline 
N_i~$&$~H_u~$&$~H_d~$&$~\theta_a~$\\ \hline \hline & & & & & & & & &  \\
$Y~$&${1/3}~$&$-{4/3}~$&${2/3}~$&$-1~$&$2~$&$0~$&$1~$&$-1~$&$~0~$\\ 
\hline & & & & & & & & &  \\
$X~$&$a^{}_i$&$b^{}_i$&$c^{}_i$&$d^{}_i$&$e^{}_i$&$f^{}_i$&$s^{}_1~$&$s^{}_2~$
&$~x_a~$\\ \hline & & & & & & & & &  \\
$Y_{k}^{}~$&$a_{[k]i}^{}$&$b_{[k]i}^{}$&$c_{[k]i}^{}$&$d_{[k]i}^{}$&$e_{[k]i}^{}$&
$f_{[k]i}^{}$
&$t_{[k]1}^{}~$&$t_{[k]2}^{}~$&$~y_{[k]a}^{}~$\\ & & & & & & & & &  \\
\hline \end{tabular} \end{center} 
\vspace{0.4cm}

We have supplemented the list of chiral superfields by an arbitrary number of 
chiral superfields $\theta_a$; they are electroweak and color singlets, 
but are assumed to have arbitrary charges under the extra Abelian 
symmetries. Some of these fields may have hidden quantum numbers as 
well.  

The mixed anomaly coefficients that involve one $X$ and two of the
same non-anomalous symmetries do not vanish, and
are given by
\begin{equation}
\begin{array}{rcl}
 C^{}_Y&=&{1\over 3}
\sum_i(a^{}_i+8b^{}_i+2c^{}_i+3d^{}_i+6e^{}_i)+s^{}_1+s^{}_2
\ ,\\
C_{[k]}^{}&=&{1\over 3}
\sum_i(a_{[k]i}^{}+8b_{[k]i}^{}+2c_{[k]i}^{}+3d_{[k]i}^{}+6e_{[k]i}^{})
+t_{[k]1}^{}+t_{[k]2}^{}\ ,\\
C^{}_{\rm weak}&=&\sum_i(3a^{}_i+d^{}_i)+s^{}_1+s^{}_2\ ,\\
C^{}_{\rm color}&=&\sum_i(2a^{}_i+b^{}_i+c^{}_i)\ .\\ & & 
\end{array}\label{eq:anom} \end{equation}
The  mixed gravitational anomaly is  the sum of the $X$ charge
times the multiplicity
\begin{equation}
C^{}_{\rm grav}=\sum_i\left[3(2a^{}_i+b^{}_i+c^{}_i)+2d^{}_i+e^{}_i+f^{}_i\right]
+2s^{}_1+2s^{}_2+\sum_ax_a\ .\label{eq:CGX}\end{equation}
None of these coefficients vanish, and are related to one another by the
Green-Schwarz constraints.

These mixed anomalies can now be used to draw inferences on the
structure of the standard model because they are all related to the  
(anomalous) charges of the electroweak invariant combinations made out of the
standard model superfields

\begin{equation}
\begin{array}{rcl}
C^{}_{\rm color}&=&X^{}_{{\cal O}^{[u]}_{ii}}+
X^{}_{{\cal O}^{[d]}_{ii}}-3X^{}_{{\cal O}^{[\mu]}_{}}\ ,
\\
C^{}_Y+C^{}_{\rm weak}-{8\over 3}C^{}_{\rm color}&=&2X^{}_{{\cal O}^{[e]}_{ii}}-
2X^{}_{{\cal O}^{[d]}_{ii}}
+2X^{}_{{\cal O}^{[\mu]}_{}}\ ,\\
C^{}_{\rm grav}&=&3C^{}_{\rm color}+X^{}_{{\cal O}^{[e]}_{ii}}+X^{}_{{\cal O}^{[\nu]}_{ii}}
+2X^{}_{{\cal O}^{[\mu]}_{}}+\sum_a x^{}_a\ ,\\ 
&&\end{array}\label{eq:anom2}\end{equation}
with the sum taken over family space. 
Since the Kac-Moody levels of the non-Abelian factors are the 
same in string theories, the Green-Schwarz mechanism requires
$C^{}_{\rm weak}=C^{}_{\rm color}$, which generates the further relation
\begin{equation}
C^{}_Y={5\over 3} X^{}_{{\cal O}^{[u]}_{ii}}-{1\over
3}X^{}_{{\cal O}^{[d]}_{ii}}+2X^{}_{{\cal O}^{[e]}_{ii}}-3X^{}_{{\cal 
O}^{[\mu]}_{}}\ .\label{eq:anom3}
\end{equation}
We can get another relation by requiring that the Weinberg 
angle assume its  canonical ($SU(5)$) value, 
$\sin^2\theta_w=3/8$. This yields $3C_Y=5C_{\rm weak}$, which enables us to 
express the excess charge of the $\mu$ term in terms of those 
of the Yukawas
\begin{equation}
X^{}_{{\cal 
O}^{[\mu]}_{}}=X^{}_{{\cal O}^{[d]}_{ii}}-X^{}_{{\cal O}^{[e]}_{ii}}\ .\label{eq:anom4}
\end{equation} 
We can now rewrite the mixed color anomaly in terms of
the excess charges of the the Yukawa operators only 
\begin{equation}
C^{}_{\rm color}=\sum_{i}\left[X^{}_{{\cal 
O}^{[u]}_{ii}}-2X^{}_{{\cal O}^{[d]}_{ii}}+3X^{}_{{\cal O}^{[e]}_{ii}}\right]
\ .\label{eq:anom5}
\end{equation}

We can also relate the mixed anomalies to the excess charges of the
off-diagonal Yukawa operators, obtaining a similar conclusion
concerning the interfamily couplings,
\begin{equation}
C^{}_{\rm color}={1\over 2}\sum_{i\ne j}\left[X^{}_{{\cal O}^{[u]}_{ij}}-2
X^{}_{{\cal O}^{[d]}_{ij}}
+3X^{}_{{\cal O}^{[e]}_{ij}}\right]\ ,
\label{eq:anom6}
\end{equation}
implying  mixing among the families. These relations enable us to draw important 
conclusions that are independent of the details of the model.  

Since $C^{}_{\rm color}$ is not zero,  the $X$ charges of some of 
the electroweak invariants do not vanish, and they  cannot appear in the 
tree level superpotential. They 
can, however,  appear as higher dimension operators, but then they will be 
suppressed by inverse powers of the cut-off, which produces hierarchy and mixing 
among the fermions masses of the standard model to satisfy 
Eqs. (\ref{eq:anom5}) and (\ref{eq:anom6})! At this stage, the argument is not specific enough to 
determine in which sector the mixing takes place. To answer that question, we need  
more detailed  models with several additional  non-anomalous phase symmetries. 
Without right-handed neutrinos, the off-diagonal charged lepton entries can be set 
to zero, and in this case, the above equation does imply mixing among the quarks.

The remaining symmetries are restricted since all of their anomaly 
coefficients must vanish. The mixed gravitational anomalies of $Y$ already 
vanishes, and that of any $Y_{k}$ must as well. The vanishing of the other 
anomalies yield relations among the $Y_k$ charge of the Yukawa operators
\begin{equation}
Y^{[k]}_{{\cal O}^{[\mu]}_{}}= {1\over 3}\left[
Y^{[k]}_{{\cal O}^{[u]}_{ii}}+
Y^{[k]}_{{\cal O}^{[d]}_{ii}}\right]=Y^{[k]}_{{\cal O}^{[d]}_{ii}}-
Y^{[k]}_{{\cal O}^{[e]}_{ii}}\ .\label{eq:hyper8}
\end{equation}
The same equations for the hypercharges are of course trivially 
satisfied. We  have the same equations for the sum of the off-diagonal elements
\begin{equation}Y^{[k]}_{{\cal O}^{[\mu]}_{}}= {1\over 3} \sum_{i\ne j}\left[
Y^{[k]}_{{\cal O}^{[u]}_{ij}}+
Y^{[k]}_{{\cal O}^{[d]}_{ij}}\right]= \sum_{i\ne 
j}\left[Y^{[k]}_{{\cal O}^{[d]}_{ij}}-
Y^{[k]}_{{\cal O}^{[e]}_{ij}}\right]\ .\label{eq:hyper9}\end{equation}

Finally, we note that we can always consider adding to $X$ any linear combination 
of $Y$ and $Y_k$. This does not affect most of the anomaly conditions except 
$\Tr(YXX)=0$ and $\Tr(Y_kXX)=0$. Thus we can choose to take 
$s^{}_1=s^{}_2$ and $t_{[k]1}=t_{[k]2}$, respectively, but then can no longer use 
these two anomaly conditions.

\section{Models}
\subsection{ Le Petit Mod\` ele}
\def\nbar{\overline N}

This is the first in a series of simple models with Abelian symmetries 
beyond the standard model. It contains one family of quarks and leptons,  
with a right-handed neutrino and see-saw mechanism, as well as one  
electroweak singlet field $\theta$. We assume one Abelian symmetry beyond 
the standard model, with its anomalies cancelled by the Green-Schwarz 
mechanism. While very elementary, this model serves as an example of our 
ideas and procedures for generating fermion mass hierarchies. The 
superpotential, which  assumes tree-level Yukawas only for the top 
quark~\cite{JS} 
and right-handed neutrino, is of the form

\begin{equation}
W={\bf Q}{\bf\overline u}H^{}_u+L{\overline N}H^{}_u+{\bf Q}{\bf\overline 
d}H^{}_d\left({\theta\over M}\right)^{n^{}_d}+L{\overline e}H^{}_d
\left({\theta\over M}\right)^{n^{}_e}+
M\nbar\hskip 2pt\nbar\left({\theta\over M}\right)^{n^{}_o}\ .\label{eq:ld} 
\end{equation}
In order to comply with the analyticity of the superpotential, 
$n^{}_e,n^{}_d$, and $n^{}_0$ must be positive 
integers. We further assume that 
the $\mu$ term is generated through the K\"ahler potential, in the 
way suggested by Giudice and Masiero~\cite{GM},
\begin{equation}
K_{GM}=H^{}_uH^{}_d\left({\overline\theta\over M}\right)^N\ .\label{eq:le} 
\end{equation}
Beyond the usual hypercharge, these couplings are 
required to be invariant under an anomalous charge $X$. This fixes $N$ to 
be a positive integer, and forbids the appearance of a $\mu$ term in the 
superpotential. The $X$ and $Y$ charges are given by 
\vspace{0.4cm}
\begin{center}
\begin{tabular}{|c|c|c|c|c|c|c|c|c|c|}
\hline & & & & & & & & &  \\
$~{\rm Charge}~$&$~{\bf Q}~$&$\overline{\bf u}~$&
$~\overline{\bf d}~$&$~L~$&$~\overline e~$&$~\overline 
N~$&$~H_u~$&$~H_d~$&$~\theta~$\\\hline \hline & & & & & & & & &\\
$Y~$&${1/3}~$&$-{4/3}~$&${2/3}~$&$-1~$&$2~$&$0~$&$1~$&$-1~$&$~0~$\\
\hline & & & & & & & & &\\
$X~$&$a$&$b$&$c$&$d$&$e$&$f$&$s^{}_1~$&$s^{}_2~$
&$~x~$\\
& & & & & & & & &  \\
\hline \end{tabular} \end{center}
\vspace{0.4cm}
This extra symmetry can be anomalous with its mixed anomalies given by
\begin{equation}
\begin{array}{rcl}C_Y&=&{1\over 3}(a+8b+2c+3d+6e)+s^{}_1+s^{}_2\ ,\cr
C^{}_{\rm weak}&=&3a+d+s^{}_1+s^{}_2\ ,\\
C^{}_{\rm color}&=&2a+b+c\ ,\\
C^{}_{\rm grav}&=&3(2a+b+c)+2d+e+f+2s^{}_1+2s^{}_2+x\\
\end{array}\label{eq:smo2} 
\end{equation}
From the  superpotential, the equations 
\begin{equation}
a+b+s=0\ ,\qquad d+f+s=0\ ,\label{eq:lf} 
\end{equation}
determine two charges, and
\begin{equation}
n_d^{}=-{{a+c+s}\over x}\ ,\qquad n^{}_e=-{{d+e+s}\over x}\ ,\qquad 
n^{}_0=-{2f\over x}\ ,\qquad N={2s\over x}\ ,\label{eq:lg} 
\end{equation}
are the positive integers which fix the dimensions of the non-renormalizable 
interactions.

The electroweak singlet field $\theta$ takes on a vacuum value 
determined by the DSW mechanism, through the vanishing of its $D$ term,
\begin{equation}
D_X=xg\vert\theta\vert^2+{g^3\over 192\pi^2}M^2_{str}C^{}_{\rm grav}\ ,\label{eq:lh} 
\end{equation}
where $g$ is the string gauge coupling, and $M^{}_{str}$ is the string 
unification scale. Clearly  $C^{}_{\rm grav}/x$ has to be negative, otherwise 
supersymmetry is broken. It follows that
\begin{equation}
\lambda\equiv{<\theta>\over 
M^{}_{str}}=\sqrt{{g^2\over 192\pi^2}{-C^{}_{\rm grav}\over x}}  \ ,\label{eq:li}
 \end{equation}
is less than one for reasonable values of the mixed gravitational 
anomaly, and can be used as an expansion parameter in the mass ratios
\begin{equation}
{m^{}_b\over m^{}_t}=\cot\beta\lambda^{n^{}_d}\ ; \qquad {m^{}_b\over 
m^{}_{\tau}}=\lambda^{n^{}_d-n^{}_e}\ ,\label{eq:lj} 
\end{equation}
where the angle $\beta$ parametrizes the ratio of the vacuum values of 
the two Higgs doublets
\begin{equation}
\tan\beta={<H^{}_u>\over<H_d^{}>}\ .\label{eq:lk} 
\end{equation}
We also have the additional relation that expresses the $\mu$ 
parameter in terms of the gravitino mass
\begin{equation}
\mu=m^{}_{3/2}\lambda^N_{}\ .\label{eq:ll} 
\end{equation}

Using the invariances of the superpotential, we can express two of the 
anomalies in terms of the integer powers
\begin{equation}
C^{}_{\rm color}=-(N+n_d^{})x\ ;\qquad C^{}_{\rm
grav}=(1-3n_d^{}-n^{}_e-2N)x
\ ,\label{eq:pmc} 
\end{equation}
so that the Kac-Moody level is given by
\begin{equation}
k_{\rm color}=12{N+n_d^{}\over 3n_d^{}+n^{}_e+2N-1}\ .\label{eq:lm} 
\end{equation}
For consistency it must be a positive integer.
Using Eq. (\ref{eq:pmc}), the non-vanishing of the color anomaly implies $N\ge
0$, or $n^{}_d\ge 0$, or both.  
Also, it tells us that, 
when $N$ and $n_d$ are positive integers,  $C_{\rm grav}/x$ is negative
and the DSW mechanism  preserves 
supersymmetry.  $X$-symmetry is broken, producing the expansion parameter
\begin{equation}
\lambda\equiv{<\theta>\over 
M^{}_{str}}=\sqrt{{g^2\over 192\pi^2}M^{}_{str}{2N+3n_d+n_e-1\over 
N+n_d}}  \ .\label{eq:ln} 
\end{equation}

In terms of anomalies, the mass ratio becomes~\cite{BR}
\begin{equation}
{m_b^{}\over m^{}_\tau}=\lambda^{(C_Y+C_{\rm weak}-{8/3}C_{\rm color})/2x-N}\ .
\label{eq:lo} 
\end{equation} 
We can express this equation in more physical terms by appealing to the Green-Schwarz mechanism,
according to which the anomalies must satisfy Eq. (\ref{eq:gs})
\begin{equation}{C^{}_{\rm grav}\over 12}={C^{}_Y\over k^{}_Y}={C^{}_{\rm 
weak}\over k^{}_{\rm weak}}
={C^{}_{\rm color}\over k^{}_{\rm color}}={C^{}_X\over k^{}_X}\ .\label{eq:lp} \end{equation}
In string compactifications, we have
$k_{\rm weak}=k_{\rm color}$, so that 
\be
C_{\rm weak}{}=C_{\rm color}^{}\ .\label{eq:weqc}
\ee
The Ib\'a\~nez relation 
\be
k_Y=\tan^2\theta_w k_{\rm weak}\ ,\qquad
C_Y^{}=\tan^2\theta_wC_{\rm weak}^{}\ ,\label{eq:iba2} 
\end{equation}
is then used to fix the Weinberg angle.
We then find that~\cite{BR,nir} 
\begin{equation}
{m_b^{}\over m^{}_\tau}={\mu\over 
m_{3/2}}\lambda^{(C_{\rm weak}/2x)(\tan^2\theta_w-5/3)}\label{eq:nice} 
\end{equation}
This equation is consistent with the data. The left-hand side is of order 
one. Numerical simulations of the MSSM with soft supersymmetry breaking 
suggest that the $\mu$ parameter is also of the order of the gravitino 
mass. In this case, this equation implies that $\tan^2\theta_w=5/3$, 
which agrees very well with the extrapolated value at unification! This 
also happens to be the value of compactifications that go through $SO(10)$. The 
agreement of this formula with experiment lends credence to our scenario 
of mass hierarchies.

We can proceed, using Eq. (\ref{eq:weqc}), to express all the charges and 
anomalies in terms of  the positive integers $N, n^{}_e,n_d^{},n^{}_0$,
\begin{equation}
a=-{x\over 6}(3N+(n_o^{}+2n_d^{}))\ ,\qquad b={x\over 
6}(n_o^{}+2n_d^{})\ ,\qquad c={x\over 6}(n_o^{}-4n_d^{})\ ,\label{eq:lq} 
\end{equation}
\begin{equation}
d={x\over 2}(n_o^{}-N)\ ,\qquad
e=-{x\over 2}(n_o^{}+2n_e^{})\ ,\qquad f=-{x\over 2}n_o^{}
\ ,\qquad s={x\over 2}N\ .\label{eq:lr} 
\end{equation}
In particular it follows that
\begin{equation}
C^{}_Y=x\left({N+n_d^{}\over 3}-2n^{}_e\right)\ .\label{eq:ls} 
\end{equation}
If we fix the Weinberg angle at its $SU(5)$ value, we can generate one 
more relation  between the integers
\begin{equation}
N+n^{}_d=n^{}_e \ .\label{eq:lt} 
\end{equation}
Hence $n_e^{}\ge n^{}_d$, and from ~\ref{eq:pmc}  $n^{}_e\ne 0$, 
resulting in a hierarchy between lepton and quark masses. 
The Kac-Moody level number is now given by
\begin{equation}
k_{\rm color}=12{n^{}_e\over 3n^{}_e+n^{}_d-1}\ .\label{eq:lu} 
\end{equation}
There is no solution with  $k_{\rm color}=1,2,3$. The lowest 
interesting value, $k_{\rm color}=4$, requires   
 $n^{}_d=1$.

In general, however we expect to have a hidden sector with non-Abelian 
gauge groups. The Green-Schwarz mechanism requires 
that the mixed anomalies with the hidden sector gauge group not 
vanish, but be proportional to $C_{\rm color}^{}$. This implies that there are 
massless fermions in the hidden sector with $X$ charges. These 
fermions will in turn contribute to the mixed gravitational anomaly. 
In the above we have not taken into account the contribution of these 
fermions. We should write instead
\begin{equation} C^{}_{\rm grav}=(C^{hid}_{}+1-3n_d^{}-n^{}_e-2N)x\ ,\label{eq:lv}
 \end{equation}
so that 
\begin{equation}
k_{\rm color}=12{N+n_d^{}\over 3n_d^{}+n^{}_e+2N-1-C^{hid}_{}}\ ,\label{eq:lw} 
\end{equation}
where 
$C^{hid}_{}x$ is the contribution from the hidden sector massless 
fermions. This shows that this contribution must be present in order 
to have a low value for the Kac-Moody integer $k_{\rm color}$. Alternatively, if
we require $k_{\rm color}=1$, as in models that do not have grand unified
groups, we find that the contribution to the mixed gravitational
anomaly from the massless particles in the hidden sector is fixed to
be
\begin{equation}C^{hid}_{}=n^{}_d-9n^{}_e-1\ .\label{eq:lx} 
\end{equation}
If the hidden sector contains a non-Abelian symmetry, its Kac-Moody level 
will also be equal to one. Thus we know from the Green-Schwarz 
cancellation that its mixed anomaly coefficient is just $-n_ex$. Thus 
these equations give us a glimpse of the physics of the hidden sector as 
well.

This instructive example serves as an illustration of the 
 power of  the Green-Schwarz restrictions. It yields the correct sign of 
the mixed gravitational anomaly. If we fix the Weinberg angle, we find 
this model to be very restrictive. First the expansion parameter is found 
to be  
\be
\lambda=\sqrt{{g^2n_e\over 192\pi^2}}  \ .\label{eq:pmlam} 
\end{equation}
Since   
\be
{m_b^{}\over m^{}_\tau}=\lambda^N={\mu\over 
m_{3/2}}\ ,
\ee
to get $m_b\sim m_\tau$~\cite{btau}, we deduce $N=0$,   the 
original proposal of Giudice and Masiero. Then
\be
{m_b^{}\over m^{}_t}=\cot\beta\lambda^{n_e}\ .\label{eq:pmres}
\ee
Since $n_e$ is a positive integer, we see that the bottom mass is 
suppressed, suggesting that $\tan\beta$ need not be that large. 
Finally we note that all the predictions of this model are, once the 
Weinberg angle is fixed,  in terms of one positive 
integer $n_e$. 

\subsection{ Le Moins Petit Mod\` ele}

We can improve the petit Mod\`ele in several ways. One is to increase the 
number of chiral families to the observed three.  Less obvious is the 
addition of pairs of fermions which are vector-like with respect to the 
standard model, but  chiral with respect to symmetries beyond. These 
fermions  appear as chiral fermions in the effective theory and 
contribute to anomalies that do not involve electroweak and color 
quantum numbers. Below the scale at which the non-standard 
symmetries break, these fermions acquire masses. A second way to 
improve on the model is to increase the gauge symmetry by adding to the 
$X$ symmetry non-Abelian or Abelian non-anomalous symmetries. 

Thus consider a standard model with one chiral family and two Abelian symmetries, 
one of which is anomalous. 
Assume two electroweak singlet fields, $\theta_a$, $a=1,2$. Its superpotential is 
\begin{equation}
\begin{array}{rcl}
W={\bf Q}{\bf\overline u}H^{}_u&+&L{\overline N}H^{}_u
+{\bf Q}{\bf\overline 
d}H^{}_d\left({\theta_1\over M}\right)^{n^{(1)}_d}
\left({\theta_2\over M}\right)^{n^{(2)}_d}\\
&+&L{\overline e}H^{}_d
\left({\theta_1\over M}\right)^{n^{(1)}_e}\left({\theta_2\over 
M}\right)^{n^{(2)}_e}+
M\overline N_{}\overline N\left({\theta_1\over 
M}\right)^{n^{(1)}_o}\left({\theta_2\over 
M}\right)^{n^{(2)}_o}\ ,\\\end{array}\label{eq:mmp} 
\end{equation}
where $n^{(a)}_e,n^{(a)}_d$, and $n^{(a)}_0$ are positive 
integers. The $\mu$ term is generated through the K\" ahler potential

\begin{equation}
K_{GM}=H^{}_uH^{}_d\left({\overline\theta_1\over 
M}\right)^{N_1}\left({\overline\theta_2\over M}\right)^{N_2}\ 
.\label{eq:mmmu} 
\end{equation}

The scales at which the extra symmetries are broken are determined by the DSW 
mechanism
\begin{equation}
x^{}_1\vert \theta^{}_1\vert^2_{}+x^{}_2\vert \theta^{}_2\vert^2_{}+{g^2\over 
192\pi^2}M^{2}_{\rm Planck}C_{\rm grav}^{}=0\ .\label{eq:mmdx} 
\end{equation}
The D term for the non-anomalous symmetry must vanish so as not to break 
supersymmetry 
\begin{equation}
y^{}_1\vert \theta^{}_1\vert^2_{}+y^{}_2\vert 
\theta^{}_2\vert^2_{}+\cdots=0\ .\label{eq:mmdo} 
\end{equation}
Thus unless the charges are very large, we expect both $X$ and $Y_1$ 
to break at similar scales. 

We note that it is possible to generate a much lower breaking scale for 
$Y_1$  by 
taking into account soft supersymmetry breaking. For instance, the 
tree-level hypercharge $D$ term does not vanish (unless $\beta=\pi/4$), 
because the soft supersymmetry breaking terms induces a negative mass 
squared. 

With only one 
chiral family, the most general non-anomalous $Y^{}_1$,  consistent 
with the tree-level superpotential,   is of the form
\be
Y_1^{}=w(B-L)+zI_{3R}+y(n_{\theta_1}-n_{\theta_2})\ ,\label{eq:mpma}
\ee
where the first two symmetries are contained within $SO(10)$. Since 
this symmetry is vector-like with respect to the $\theta$ fields, the  
vanishing of its D-term requires 
\be
<\vert\theta_1\vert>=<\vert\theta_2\vert>\equiv <\vert\theta\vert>\ .\label{eq:theta}
\ee
Hence both $X$ and $Y_1$ are broken at the same scale. Invariance under $Y_1$ yields
\be
n^{(1)}_{d,e}=n^{(2)}_{d,e}\equiv n^{}_{d,e}\ ,\label{eq:two}
\ee
and  there is only one expansion parameter.
This case is not very different from the petit Mod\`ele. We find that 
\be
C^{}_{\rm grav}=(1-3n_d-n_e-2N)(x_1+x_2)\ ,
\ee
which has the proper sign to preserve supersymmetry. Here $N=N_1=N_2$. 
however we are left with much arbitrariness. 

In special cases, this model has some distinctive features. For instance 
if $Y_1$ is chosen so that  $y=0$, it can be broken at a 
very different scale, say when the right-handed sneutrino gets a vacuum 
value, perhaps around $10^{11}$ GeV. 
Another choice is when $z=0$, in which case, the $X$ symmetry loses its 
anomaly, contradicting our assumptions. Finally, when $w=0$, we find that 
$n_d=0$ indicating no suppression of the bottom to top ratio.
Thus unless the non-anomalous symmetry is determined in a special way, we 
do not have enough data to narrow down the anomaly structure. With 
several families, this situation may be different. We leave to a future 
publication~\cite{ir} the study of such 
models.  
\section*{Acknowledgments}I wish to thank the organizers, H. Itoyama, M. 
Kaku, H. Kunimoto, M. Ninomiya, and H. Shirokura  for their 
marvelous hospitality. This work is supported in part by the United 
States Department of Energy under grant DE-FG05-86-ER40272.


\begin{thebibliography}{Ref}
\bibitem{FN} C.~Froggatt and H.~B.~Nielsen Nucl. Phys. B147 (1979) 277.
\bibitem{BR} P. Bin\'etruy and P. Ramond, Phys. Lett. B350 (1995) 49.
\bibitem{BLR} P. Bin\'etruy, S. Lavignac and P. Ramond, preprint 
LPTHE-ORSAY 95/54, UFTIFT-HEP-96-1, hep-ph/9601243.
\bibitem{GS} M. Green and J. Schwarz, Phys. Lett. B149 (1984) 117.
\bibitem{Ib} L. Ib\'a\~nez, Phys. Lett. B303 (1993) 55.
\bibitem{IR} L. Ib\'a\~nez and G. G. Ross, Phys. Lett. B332 (1994) 100.
\bibitem{top} CDF collaboration, F. Abe {\em et al.}, Phys. Rev. Lett. 74
(1995) 2626; D0 collaboration, S. Abachi {\em et al.}, Phys. Rev. Lett. 74
(1995) 2632.
\bibitem{RRR}P. Ramond, R.G. Roberts and G.G. Ross, Nucl. Phys. B406 (1993) 19.
\bibitem{wolf} L. Wolfenstein, Phys. Rev. Lett. 51 (1983) 1945.
\bibitem{faraggi} A.E. Faraggi, Phys. Lett. B274 (1992) 47, Phys. Rev. D47
(1993) 5021; A.E. Faraggi and E. Halyo, Phys. Lett. B307 (1993) 305, Nucl.
Phys. B416 (1994) 63.
\bibitem{DSW} M. Dine, N. Seiberg and E. Witten, Nucl. Phys. B289 (1987) 317;
J. Atick, L. Dixon and A. Sen, Nucl. Phys. B292 (1987) 109.
\bibitem{ir} N. Irges and P. Ramond, in preparation.
\bibitem{JS} V. Jain and R. Shrock, Phys. Lett. B352 (1995) 83;  
preprint ITP-SB-95-22 (hep-ph/9507238).
\bibitem{GM} G.F. Giudice and A. Masiero, Phys. Lett. B206 (1988) 480.
\bibitem{btau} H.~Arason, D.~J.~Casta\~no, B.~Keszthelyi, S.~Mikaelian,
E.~J.~Piard, P.~Ramond, and B.~D.~Wright, Phys. Rev. Lett. 67 (1991) 2933;
A.~Giveon, L.~J.~Hall, and U.~Sarid, Phys. Lett. 271B (1991) 138.
\bibitem{nir} Y. Nir, Phys. Lett. B354 (1995) 107.

















\end{thebibliography}
\end{document}